# Pressure-induced symmetry lowering in Nb$_3$Sn$_{1-x}$ superconductor

V. Svitlyk*, M. Mezouar

*ID27 High Pressure Beamline, European Synchrotron Radiation Facility, Grenoble, France*

* svitlyk@esrf.fr

**Abstract**

Cubic *Pm*-3*n* Nb$_3$Sn$_{0.92}$ superconductor ($T_c$ ~ 16 K) was found to exhibit tetragonal instabilities at the superconducting state ($T$ = 10 K). These instabilities are manifested through the appearance of reflections which are forbidden in the *Pm*-3*n* symmetry but are compatible with the *P*4$_2$/*mmc* structure which is observed in the Nb$_3$Sn$_{1-x}$ system for higher Sn content at temperatures lower than ~ 43 K. Nevertheless, the low-temperature structure of Nb$_3$Sn$_{0.92}$ remains metrically fully cubic, as concluded from single crystal synchrotron radiation diffraction experiments. Subsequent application of external pressure amplifies the observed instabilities with a resulting pseudo-cubic – tetragonal transformation at $P$ = 3 GPa at 10 K and this transition is energy driven, as concluded from *ab initio* calculations. The electronic structures of the corresponding phases are virtually identical and, therefore, the pseudo-cubic – tetragonal transformation does not influence significantly the underlying electronic interactions. Consequently, no anomalies in the behavior of the critical temperature, $T_c$, are expected at this pressure. However, anomalies in the upper critical field are anticipated during this transition, in analogy to the corresponding behavior observed during the cubic-tetragonal transformation in Nb$_3$Sn$_{1-x}$ induced by increase in Sn content. Therefore targeted changes in composition could be used to enhance upper critical field of Nb$_3$Sn$_{1-x}$ for specific extreme conditions of temperature and pressure.

## 1. Introduction

Superconductivity in Nb$_3$Sn was discovered in 1954 with a reported onset of superconducting critical temperature, $T_c$, of 18 K [1]. Due to a remarkably high critical current densities of more than 1300 A/mm$^2$ at 15 T [2] this phase is currently a primary target material for applications in state-of-the-art accelerator and plasma-control systems [3–5]. As a result considerable effort is made by scientific community to further enhance performance of Nb$_3$Sn and one of the key ways is the employment of artificial pinning centers [6]. The latter can be introduced by selective compositional doping [7–9] and, therefore, control of exact stoichiometry is important for this system.

The Nb$_3$Sn phase exhibits a rich panel of structural properties which, in turn, influence the corresponding physical behavior. Structure of this phase belongs to the A15 series of compounds (*Pm*-3*n*) and is composed of edge-shared SnNb$_{12}$ icosahedra. The A15 lattice yields shorter distances between the Nb atoms compared to that of pure bcc Nb metal (2.65 [10] vs. 2.86 Å [11]). This results in an enhanced overlap between the Nb *d* orbitals in Nb$_3$Sn which generates narrower and higher electronic density of states at the Fermi level [12]. This, in turn, yields doubling of the superconducting transition temperature in Nb$_3$Sn (~ 18 vs. ~9 K in bcc Nb metal [13]).

The Nb$_3$Sn system features region of compositional homogeneity due to possible deficiency of Sn sites with a resulting stoichiometry range of $0.18 \leq \beta \leq 0.25$ (Nb$_{1-\beta}$Sn$_\beta$ compositional representation) [14,15]. In addition, below $T_M$ of ~ 43 K a tetragonal $P4_2/mmc$ phase of Nb$_3$Sn was found to exist in a rather narrow range of $\beta$ ($0.245 < \beta < 0.252$) and this symmetry lowering results from a shear transformation in the parent cubic phase [12,16,17]. The cubic-tetragonal phase transition triggers anomalies in corresponding physical properties, namely resistance and specific heat [18]. Thus these anomalies can be used to track the temperature of the cubic-tetragonal structural transition, $T_M$, under extreme conditions of pressure through measurements of the corresponding physical response.

The $T_c$ is reported to vary rather smoothly as a function of $\beta$ [14][19] and its behavior can be empirically modelled with Boltzman function [12]:

$$T_C(\beta) = \frac{-12.3}{1+\exp(\frac{\beta-0.22}{0.009})} + 18.3 \quad \text{Eq. 1.}$$

Since no apparent anomaly in $T_c$ is observed at the cubic-tetragonal phase boundary point ($\beta = 0.245$) the differences in corresponding electronic structures should be rather negligible with no resulting significant impact on the underlying electron-electron coupling. However, anomaly in the upper critical field is observed around this transition point [12] which should be taken into account for application purposes.

Along with compositional tuning high pressure (HP) is yet another mean to influence the observed structural and physical properties of this phase. Indeed, application of HP induces increase in the temperature of the cubic-tetragonal transformation, $T_M$ [20]. Contrary, application of HP induces decrease in $T_c$ of Nb$_3$Sn [18,20,21] with a $dT_C/dT_P$ rate of -1.4x10$^{-2}$ K kbar$^{-1}$. Opposite pressure effects on the $T_M$ and $T_c$ values can be explained by the pressure-induced charge redistribution, in particular of the *d*-electron bands, and this phenomenon can be qualitatively described with the Weger-Labbe-Friedel model [22–24]. This system, is therefore, features very intimate correlations between structural and electronic properties. Nevertheless, no details of structural evolution under superconducting condition as a function of external pressure have been reported in the literature. Establishing of the corresponding structural behavior is crucial for understanding of the underlying composition-structure-properties relationships which, in turn, can allow more accurate modeling of the observed superconducting properties. In this paper we report results of detailed HP single crystal diffraction experiments on Nb$_3$Sn at ambient and superconducting temperatures which allowed to discover new fine structural states in this compound and to revisit the corresponding low-temperature (LT) – HP phase diagram. Experimental results are complemented with *ab initio* calculations in order to get more insight into the underlying energy contributions.

## 2. Experiment

Two single crystals of Nb$_3$Sn with a typical size of 20x20x20 µm$^3$ were extracted from a commercial sample (Goodfellow, 99.9% starting purity of raw metals). Initially, the first Nb$_3$Sn crystal was compressed at room-temperature (RT) up to 7 GPa in order to obtain reference data for the subsequent LT measurements. For RT studies the crystal was loaded in a diamond anvil cell (DAC) with 600 µm culets. Rhenium gasket with a 300 µm hole and initial pre-indented thickness of 90 µm was used as a sample chamber. Second crystal was compressed at 10 K up to 11 GPa in order to study structural and phase behavior of Nb$_3$Sn in superconducting regime. For this experiment crystal was loaded in a LT-DAC (600 µm culets) and contained in a stainless-steel gasket (300 µm hole, 95 µm thickness). In-house developed He-flow cryostat was used to achieve and control LT. For both experiments He gas was used as a pressure-transmitting medium since it preserves excellent hydrostaticity up to at least 50 GPa [25]. Ruby spheres loaded together with the crystals were used to monitor pressure using ruby fluorescence technique [26]. Typical pressure steps for experiments at RT and 10 K were of 1.5 GPa. No bridging, i.e. sample clamping between opposite diamonds, was observed for the both pressure ramps.

Angle-resolved single crystal diffraction data were collected at the ID27 High Pressure Beamline at the European Synchrotron Radiation Facility (ESRF, Grenoble, France). The wavelength of synchrotron radiation was set to 0.3738 Å and beam was focused down to 3x3 µm$^2$. During data collection DACs were oscillated in a 70º range with a 1º step and raw data were recorded on a Mar165 CCD detector. Subsequently data were reduced and experimental slices of the reciprocal space generated with CrysAlisPro suite [27]. Crystal structures were refined with SHELXL software [28]. Simulation of theoretical slices of the reciprocal space and determination of primary order parameters for structural transformations were done using the ISODISTORT package [29,30].

*Ab initio* calculations have been performed for cubic and tetragonal modifications of $Nb_3Sn$ in order to elucidate electronic contributions to the corresponding transformation. Calculations were performed with a Wien2k 18.2 package [31] using the Perdew-Burke-Ernzerhof generalized gradient approximation exchange correlation potential [32]. Density of states (DOS) were calculated for the *Pm-3n* and *P4$_2$/mmc* $Nb_3Sn_{0.92}$ structures observed at pressures of 0.16 and 3.21 GPa, which corresponds to the experimental data presented in Tables 1 and 2, respectively. For calculations the Brillouin zone was divided with a *k* mesh of 1000 (10x10x10) points and Sn sites were considered as fully occupied.

## 3. Results and discussion

### 3.1. Crystal structure and phase transition

Both crystals were found to be deficient in Sn, as concluded from single crystal data analysis. The corresponding occupancies on Sn sites were refined as a function of pressure for all the data points and were found to remain constant within an error range for two samples. The resulting compositions, which correspond to the averaged occupancies, are $Nb_3Sn_{0.90(2)}$ ($\beta$ = 0.231 in $Nb_{1-\beta}Sn_\beta$ representation, crystal used for RT compression) and $Nb_3Sn_{0.92(1)}$ ($\beta$ = 0.235, LT compression at 10 K). The corresponding $T_c$s were obtained from the Eq. 1 and are equal to 15.5 and 16.4 K for crystals studied at RT and LT, respectively. The values of $\beta$s which are smaller than 0.245 indicate that both crystals should adapt cubic *Pm-3n* symmetry at all the temperature range, i.e. no LT transitions to tetragonal structures should occur. Indeed, RT single crystal diffraction data of the $Nb_3Sn_{0.90}$ sample are fully compatible with the expected cubic *Pm-3n* structural arrangement up to the highest studied pressure of 7 GPa. Similarly, at RT and ambient pressure the $Nb_3Sn_{0.92}$ crystal exhibits only features expected for the cubic $Nb_3Sn$ structure (Fig. 1, left). This symmetry was also preserved after loading of the DAC with He gas ($P_0$ = 0.2 GPa) for a further LT experiment.

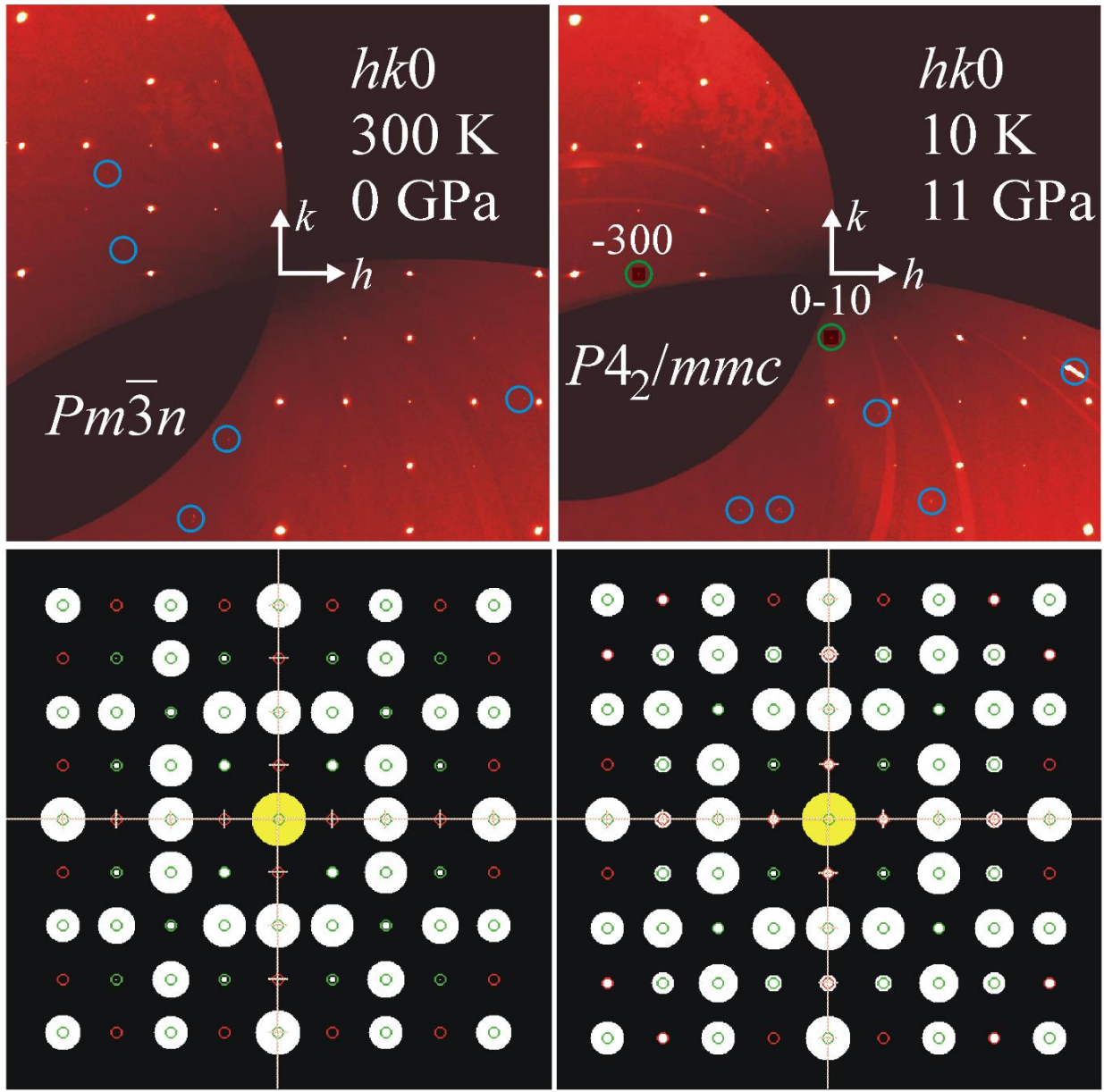

Figure 1. Experimental (top) and simulated (bottom) $hk0$ slices of reciprocal space of $Nb_3Sn_{0.92}$ at ambient conditions (left) and 10 K and 11 GPa (right). On the simulated slices (bottom) green and red open circles correspond to symmetry allowed and forbidden reflections, respectively, in the cubic $Pm\text{-}3n$ setting. Relative intensities of the observed reflections are proportional to the area of the corresponding filled white circles. Simulated transition to the tetragonal symmetry (bottom, right) generates forbidden in the cubic phase reflections (e.g. observed -300 and 0-10 marked by green circles on the top, right) which are allowed in the $P4_2/mmc$ structure. Contrast around these reflections was increased for visualization purposes (top, right). Blue circles (top) mark reflections not commensurate with the reciprocal space of $Nb_3Sn_{0.92}$ and which originate from sample environment (diamonds, ruby spheres).

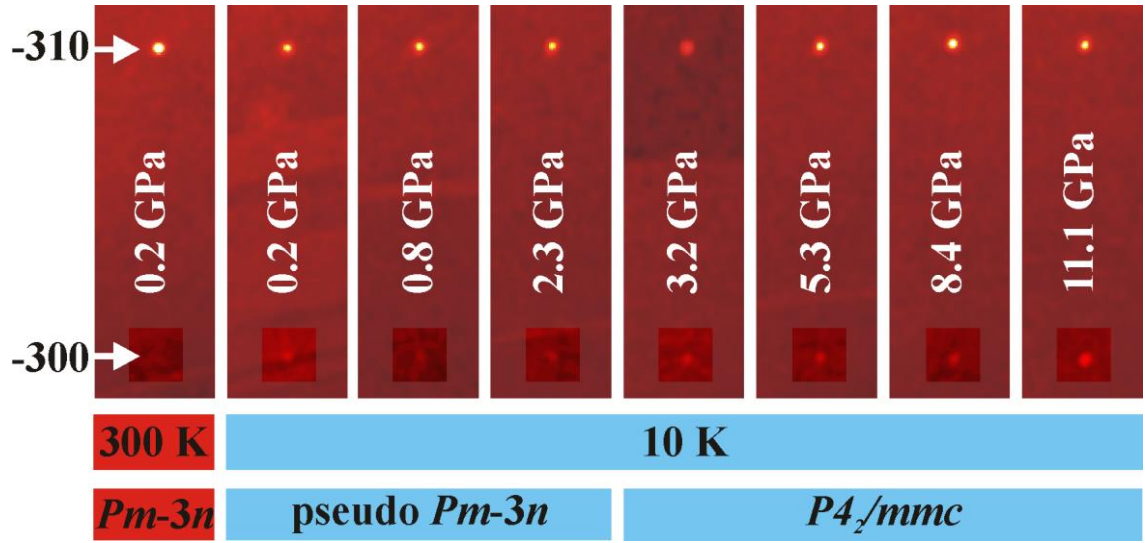

Figure 2. Emergence of -300 reflection in $Nb_3Sn_{0.92}$ upon cooling and its evolution with pressure. -310 reflection is shown as a reference. Contrast around the -300 reflection was increased for visualization purposes.

Surprisingly, weak reflections which violate the cubic *Pm-3n* symmetry appeared upon cooling of the $Nb_3Sn_{0.92}$ crystal down to 10 K (Fig. 1, top, right, -300 and 0-10 reflections marked with green circles, $P$ = 11 GPa; Fig. 2, -300 reflection, $P$ = 0.2 – 11 GPa). Observed -300 and 0-10 reflections are compatible with the tetragonal $Nb_3Sn$ structure and, subsequently, could indicate onset of the *Pm-3n* - *P4$_2$/mmc* transformation in $Nb_3Sn_{0.92}$ ($\beta$ = 0.235). This is rather surprising since the tetragonal $Nb_3Sn$ structure is expected to form upon cooling in the $0.245 < \beta < 0.252$ compositional range. Indeed, freely refined *a*, *b* and *c* unit-cell parameters of $Nb_3Sn_{0.92}$ did not exhibit tetragonal splitting at 10 K and 0.2 GPa. Moreover, least-square refinement in a tetragonal model (split Nb site) using data collected at these conditions yielded a model equivalent to the cubic *Pm-3n* structure within an error range. Weak intensity of the emerged *P4$_2$/mmc* reflections could explain observed equivalence of the obtained tetragonal structure to its parent cubic phase. However, intensity of the -300 reflection (Fig. 2) remains constant within the error range up to 11 GPa (Fig. 3, left), thus indicating that the effective transition could occur around this pressure. Surprisingly, the *a*, *b* and *c* parameters show clear and sudden divergence to a tetragonal symmetry already at 3 GPa (Fig. 3, right, Fig. 4, left). In addition, the Nb2 atom decouples from the Nb1 atom also at the pressure of 3 GPa, as seen by the divergence of its *x* coordinate from the original $x = 1/4$ value (Fig. 4, right). This pressure is, therefore, to be considered as a formation point of a tetragonal $Nb_3Sn_{0.92}$ at 10 K. The $Nb_3Sn_{0.92}$ structure at pressures below 3 GPa at 10 K is metrically fully cubic and, therefore, will be denoted as pseudo-cubic. Summary of structural refinement of single crystal data of $Nb_3Sn_{0.92}$ at 0.16 GPa (cubic symmetry) and 3.21 GPa (tetragonal symmetry) are presented in Tables 1 and 2, respectively.

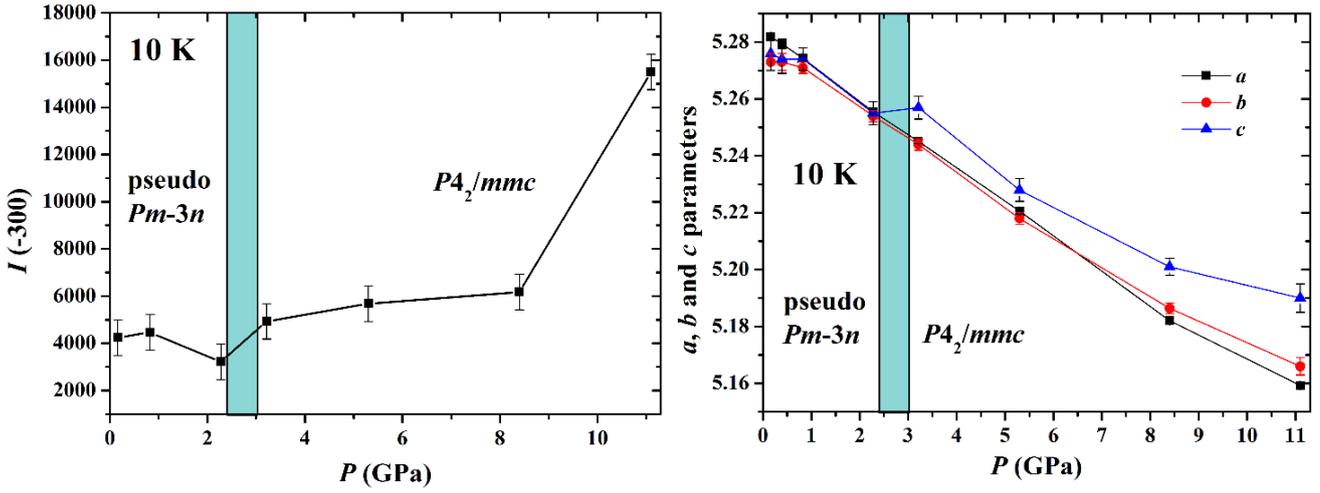

Figure 3. Intensity of -300 reflection of $Nb_3Sn_{0.92}$ as a function of pressure at 10 K (left); divergence in *a*, *b* and *c* unit-cell parameters at 3 GPa indicating onset of a metrically tetragonal structure around this pressure.

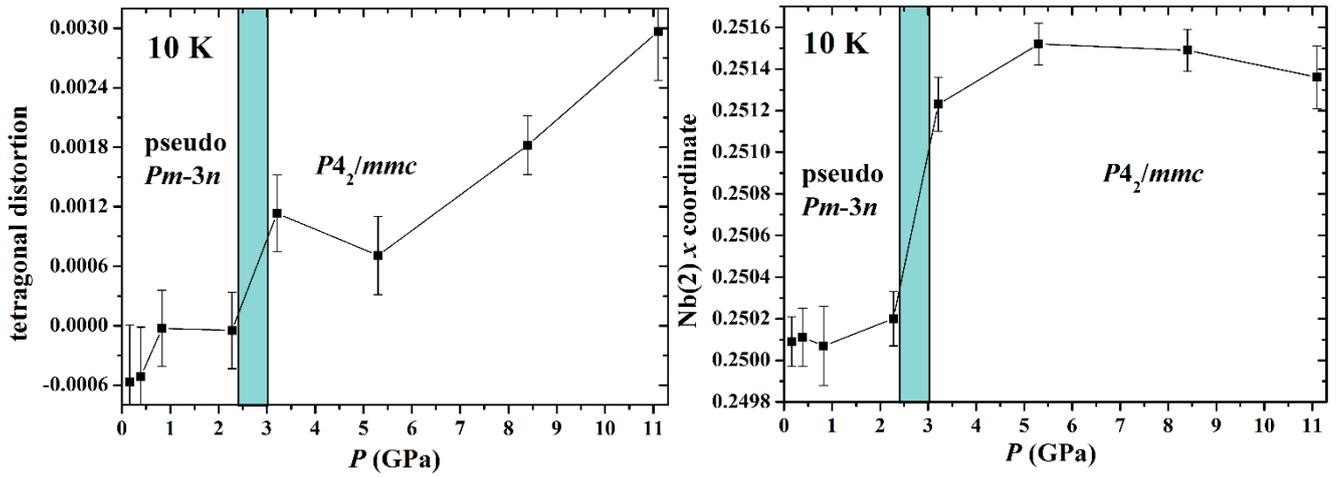

Figure 4. Tetragonal distortion of $Nb_3Sn_{0.92}$ defined as $\delta = (a-c)/(a+c)$ (left) and *x* coordinate of Nb2 refined in tetragonal symmetry as a function of pressure at 10 K (right).

Table 1. Results of structural refinement and crystal data of $Nb_3Sn_{0.92}$ at 0.16 GPa and 10 K ($Pm\text{-}3n$, $a = b = c = 5.273(2)$ Å, $z = 2$, $R_1 = 0.0524$ for reflections with $I > 2\sigma(I)$; $U^{11} = U^{22} = U^{33}$, $U^{23} = U^{13} = U^{12} = 0$).

| Atom | Site | Occ. | x | y | z | $U_{eq}$, Å$^2$ | $U^{11}$ |
|------|------|------|---|---|---|-----------------|----------|
| Nb | 6c | 1 | ¼ | 0 | ½ | 0.007(4) | 0.007(4) |
| Sn | 2a | 0.922(15) | 0 | 0 | 0 | 0.009(3) | 0.009(3) |

Table 2. Results of structural refinement and crystal data of $Nb_3Sn_{0.92}$ at 3.21 GPa and 10 K ($P4_2/mmc$, $a = b = 5.2445(13)$ Å, $c = 5.257(4)$ Å, $z = 2$, $R_1 = 0.0523$ for reflections with $I > 2\sigma(I)$; $U^{11} = U^{22}$; $U^{23} = U^{13} = U^{12} = 0$).

| Atom | Site | Occ. | x | y | z | $U_{eq}$, Å$^2$ | $U^{11}$ | $U^{33}$ |
|---|---|---|---|---|---|---|---|---|
| Nb1 | 2f | 1 | ½ | ½ | ¼ | 0.006(2) | 0.009(2) | 0.002(4) |
| Nb2 | 4l | 1 | 0.2512(1) | 0 | ½ | 0.006(2) | 0.009(2) | 0.001(4) |
| Sn | 2d | 0.932(10) | 0 | ½ | ½ | 0.009(2) | 0.011(2) | 0.004(3) |

During the cubic-tetragonal symmetry change the original Nb position is split into two sites thus providing additional degree of freedom for structural relaxation and the resulting transition is manifested as antiparallel movements of Nb atoms in the *ab* planes (Fig. 5, red arrows). The tetragonal $P4_2/mmc$ structure forms a subgroup of the parent cubic $Pm$-$3n$ modification and a corresponding Landau expansion of free energy contains invariants of a third order. Therefore the observed cubic-tetragonal symmetry lowering is a first-order structural transition of a displacive type. Indeed, despite that the absolute changes in structural parameters at the transition point ($P = 3$ GPa) are rather small they are abrupt and sudden (Figs. 3, right; 4, right) which is characteristic of first-order transformations.

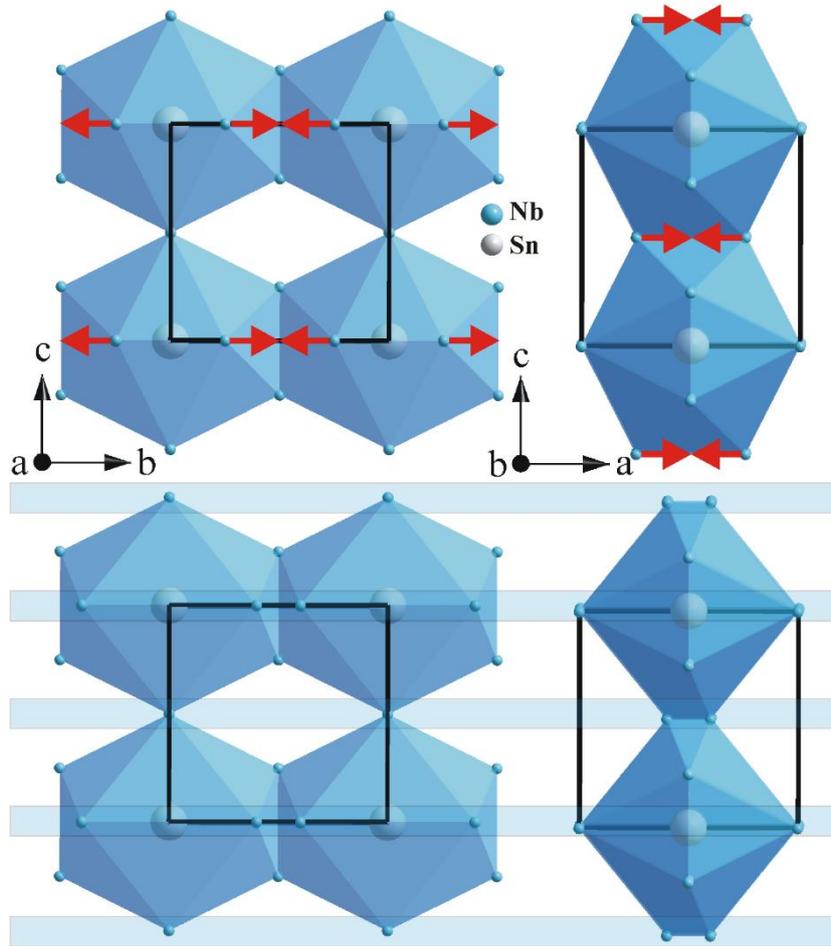

Figure 5. Anti-parallel movements of the Nb atoms during the $Pm$-$3n$ - $P4_2/mmc$ transition along the *b* (top, left) and *a* (top, right) directions; illustration of the atomic planes which are affected by the tetragonal distortion (bottom, light blue areas).

Due to a small magnitude of structural changes at the cubic-tetragonal transition point no anomaly is observed in the corresponding unit-cell volume of $Nb_3Sn_{0.92}$ within the error range (Fig. 6, left). The compression data were fitted with a $3^{rd}$ order Birch-Murnaghan equation of state (EOS):

$$P(V) = \frac{3B_0}{2}\left[\left(\frac{V_0}{V}\right)^{\frac{7}{3}} - \left(\frac{V_0}{V}\right)^{\frac{5}{3}}\right]\left[1 + \frac{3}{4}(B_0' - 4)\left\{\left(\frac{V_0}{V}\right)^{\frac{2}{3}} - 1\right\}\right] \quad \text{Eq. 2.}$$

where $V_0$ is the zero pressure volume, $B_0$ is the bulk modulus and $B_0'$ is its first pressure derivative. The tetragonal $Nb_3Sn_{0.92}$ phase was found to be slightly softer ($B_0 = 121(10)$ GPa) than the corresponding cubic modification at RT ($B_0 = 148(12)$ GPa, Fig. 6, right). The corresponding coefficients of Eq. 2 are: $V_0 = 147.9(5)$ Å$^3$, $B_0 = 121(10)$ GPa, $B_0' = 9.2$ GPa (fixed) for tetragonal $Nb_3Sn_{0.92}$ at 10 K; $V_0 = 147.6(2)$ Å$^3$, $B_0 = 148(12)$ GPa, $B_0' = 9.2$ GPa (fixed) for cubic $Nb_3Sn_{0.90}$ at RT. The $B_0'$ coefficients could not be refined independently due to a limited number of datapoints and their values were fixed to 9.2 for both phases for the final refinement cycles.

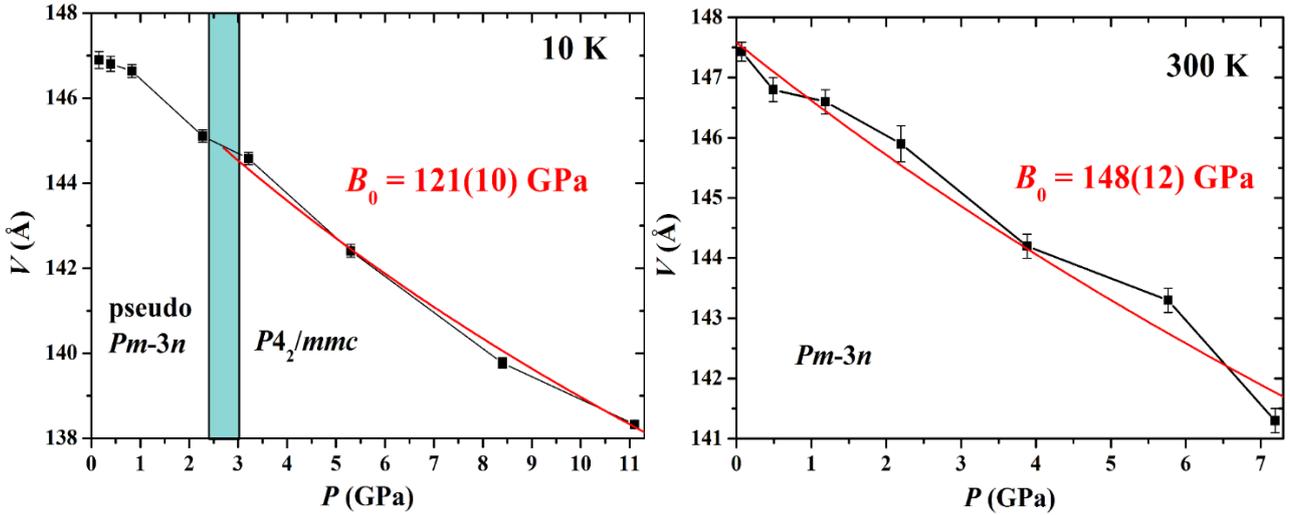

Figure 6. $P$-dependent evolution of unit-cell volume of $Nb_3Sn$ at 10 K (left) and RT (right). Red solid lines correspond to least-square fits with a $3^{rd}$ order Birch-Murnaghan EOS (Eq. 2).

### 3.2. Electronic structure and energy contributions

The observed $P$-dependent structural changes in $Nb_3Sn_{0.92}$ at 10 K and 3 GPa are rather subtle and the corresponding phases are expected to feature similar nature of electronic interactions. Indeed, the DOS graphs for the corresponding cubic and tetragonal modifications are virtually identical (Fig. 7, left, DOS of cubic $Nb_3Sn$ is shown as an example) and, as expected, exhibit strong metallic characters. Fermi level ($E_f = 0$ eV) is filled essentially by Nb $d$ electrons (Fig. 7, left, Nb states are marked by solid green line) with a contribution from electrons of mixed Sn $p$-$d$ states (Fig. 7, left, Sn states are marked by solid blue line). *Ab initio* compression was further performed in order to obtain energy evolution for both phases as a function of pressure. For this volumes of the unit-cells were varied progressively by 0.1% assuming isotropic structural response, i.e. constant *a*:*b*:*c* ratios. This assumption is valid for tetragonal Nb$_3$Sn up to c.a. 8 GPa (Fig. 3, right). Starting points for these calculations were structural models experimentally observed at 0.16 and 3.21 GPa (data in Tables 1 and 2, respectively). Every probed data point (Fig. 7, right, black and red points) corresponds to a full cycle of DFT calculations which yields a total energy of a given system (projected on the ordinate of the Fig. 7, right). This includes electron-electron repulsion, nuclear-electron attraction and exchange-correlation energies, kinetic energy of non-interacting particles and the repulsive Coulomb energy of fixed nuclei. As expected, compression increases total energy of the low-pressure (LP) cubic Nb$_3$Sn modification (Fig. 7, right, black dots) and in the vicinity of the transition point it becomes higher than the corresponding energy of the HP tetragonal phase (Fig. 7, right, red dots). At this stage the *Pm-3n* - *P4$_2$/mmc* structural transition is triggered which results in an energy reduction of the system.

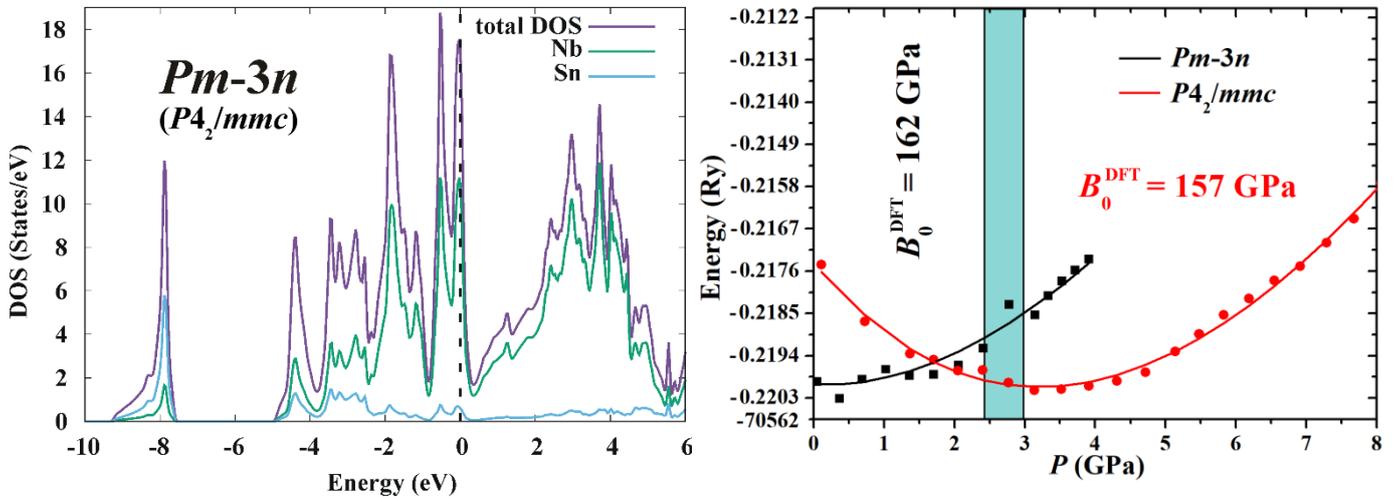

Figure 7. DOS of cubic Nb$_3$Sn (left; Fermi level, $E_f$, is marked by black dashed line; DOS of the tetragonal *P4$_2$/mmc* modification is virtually identical to the DOS of the cubic Nb$_3$Sn phase and is, therefore, not presented in a separate figure); *ab initio* compressibility of *Pm-3n* and *P4$_2$/mmc* structures of Nb$_3$Sn (right, black and red dots, respectively; solid lines represent corresponding polynomial fits for representation purposes).

For the cubic Nb$_3$Sn modification *ab initio* calculations yield results almost identical to the experimental within one standard deviation: $B_0^{DFT}$ = 162 GPa vs. $B_0^{EXP}$ = 148(12) GPa. The corresponding value of the bulk modulus is, however, somewhat overestimated for tetragonal Nb$_3$Sn: $B_0^{DFT}$ = 157 GPa vs. $B_0^{EXP}$ = 121(10) GPa. As a result the cubic-tetragonal transition pressure predicted from the *ab initio* calculations slightly differs from the one observed experimentally since the pressure scale on the Fig. 7 (right, abscissa) is based on the theoretical EOSs. The corresponding calculations have been performed at 0 K and this, in part, could be in the origin of the observed differences.

### 3.3. Phase diagram

From the presented structural data the following *T-P* phase diagram of Nb$_3$Sn$_{1-x}$ ($x$ = 0.1 for RT compression and $x$ = 0.08 for compression at 10 K) can be constructed (Fig. 8). At RT and up to at least 7 GPa cubic *Pm-3n* structure of Nb$_3$Sn$_{0.90}$ is stable. The Nb$_3$Sn$_{0.92}$ phase also exhibits *Pm-3n* symmetry at RT and 0.2 GPa (HP DAC loading pressure). However, reflections not compatible with cubic *Pm-3n* symmetry are present for this phase at 10 K but the corresponding structure remains metrically cubic (this phase is denoted as pseudo-cubic Nb$_3$Sn$_{1-x}$). The LT data clearly show that the Nb$_3$Sn$_{0.92}$ phase with a projected $T_c$ of 16.4 K features tetragonal instabilities at the superconducting regime ($T$ = 10 K). Subsequent application of pressure amplifies these instabilities and at the critical pressure of 3 GPa Nb$_3$Sn$_{1-x}$ system relaxes in to the tetragonal *P4$_2$/mmc* symmetry with a resulting decrease in the total energy, as concluded from the corresponding *ab initio* electronic structure calculations.

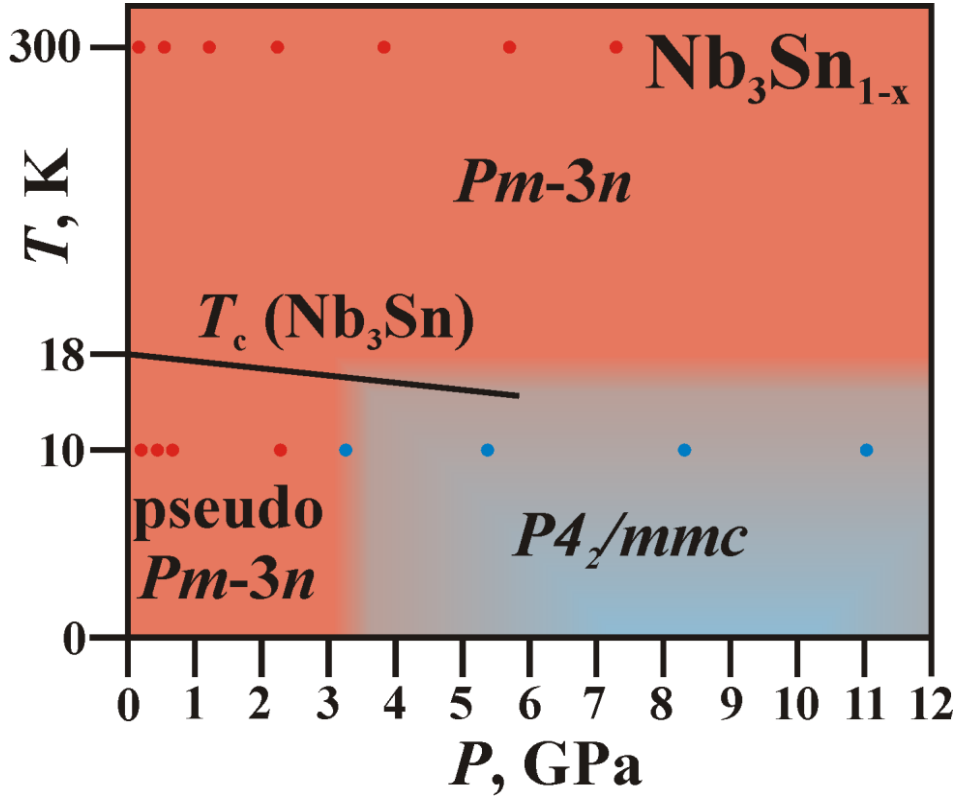

Figure 8. Phase diagram of Nb$_3$Sn$_{1-x}$ in a *T-P* space. Filled circles correspond to experimental data points. Black solid line originating at 18 K schematically separates non-superconducting and superconducting states of Nb$_3$Sn$_{1-x}$; negative ramp highlights decrease in $T_c$ upon application of HP. Light blue area corresponds to the domain of existence of the tetragonal Nb$_3$Sn$_{1-x}$ structure; in this work upper temperature boundaries were not studied for this region.

For the Nb$_3$Sn$_{1-x}$ system different compositions are expected to yield different behavior as a function of *T* and *P*. Specifically, phases with Sn content higher than 0.92, i.e. closer to a critical value of $\beta$ = 0.245 for which tetragonal *P4$_2$/mmc* structure is formed directly on cooling, could feature lower pseudo-cubic – tetragonal transition pressure ($P$ < 3 GPa). Contrary, phases poorer in Sn could require higher pressures ($P$ > 3 GPa) to induce the corresponding transformation. Finally, it can be reasonably assumed that chemical doping could be also used to change the observed behavior both a function of temperature and pressure in order to optimize superconductive performance of Nb$_3$Sn, notably values of the upper critical field.

## 4. Conclusions

Cubic $Pm$-$3n$ modification of $Nb_3Sn_{0.92}$ ($\beta$ = 0.235) was found to exhibit tetragonal instabilities under superconducting conditions, although the corresponding crystal structure remains metrically cubic. At 10 K this pseudo-cubic phase undergoes a $P$-induced symmetry lowering towards a tetragonal $P4_2/mmc$ structure at $P$ = 3 GPa. Electronic structures of the corresponding cubic and tetragonal modifications are virtually identical, as concluded from *ab initio* calculations based on experimental data, which indicates that this transition should not influence underlying electronic interactions. This is in agreement with the observed superconductive behavior in the $Nb_3Sn_{1-x}$ system which does not show anomalies in $T_c$ upon the $Pm$-$3n$ – $P4_2/mmc$ transition as a function of $x$. The observed pseudo-cubic – tetragonal transition in $Nb_3Sn_{0.92}$ was found to be energy driven, i.e. corresponding structural relaxation as a function of $P$ results in lowering of total energy of the system. No deviations from the cubic $Pm$-$3n$ symmetry were observed at RT for the studied $Nb_3Sn_{1-x}$ samples.

Observed rich structural behavior in the $Nb_3Sn_{1-x}$ system opens additional ways to influence corresponding superconductive properties. Namely, variable stoichiometry / composition could be used to tailor response of this system, e.g. values of the upper critical field, for specific extreme conditions of temperature and pressure. At the same time this work calls for further research on the $Nb_3Sn_{1-x}$ system for a larger range of $x$ and in a wider $T$-$P$ domain. Notably, more information on the HT cubic – LT pseudo-cubic transition is needed to complement the existing phase diagram and to establish possible correlations with a cubic-tetragonal transition observed in this system around 43 K for a higher Sn content.